# Propagation of pulse fluctuations of multimode sources in optical fibers and waveguides with arbitrary chromatic dispersion

**José Capmany***

*ITEAM Research Institute, Universitat Politècnica de València, Camino de Vera s/n, 46022 Valencia, Spain*
*jcapmany@iteam.upv.es

**Abstract:** We extend Marcuse's analysis of pulse fluctuations of randomly phased multimode sources to arbitrary chromatic dispersion, chirp, finite mode linewidth, mode-partition noise and finite receiver bandwidth. On long fibers the fluctuations depend only on the shape of the group-delay curve: a monotonic delay separates the modes, whereas a symmetric delay leaves pairs of modes with identical intensity profiles and a persistent beat noise of up to $1/\sqrt{2}$. Dispersion down-converts the beats of overlapping modes to baseband, where receivers cannot filter them. The detected power follows random-phasor statistics, and Gaussian estimates of fade probability can be wrong by orders of magnitude. Mode-partition noise grows with dispersion and dominates in band-limited links.

## 1. Introduction

Between 1980 and 1981 Marcuse published a series of papers that set the framework for the analysis of linear pulse propagation in single-mode fibers [1–3]. A Gaussian power pulse modulating a source of Gaussian spectrum was propagated through a fiber whose propagation constant was expanded up to its third derivative, that is, including first-order (group-velocity) and second-order dispersion in Marcuse's terminology. Closed-form expressions were derived for the spectrum and the rms width of the ensemble-averaged pulse, first for a single-line source [1], then for a source with several discrete spectral lines of arbitrary power and possibly skewed envelope [2], and finally for chirped sources [3]. In 2003 this formalism was extended to an arbitrary number of dispersion coefficients, including source chirp and linewidth [4], which allows the analysis of media with complex dispersion profiles such as dispersion-flattened and photonic-crystal fibers [5] and connects naturally with temporal imaging [6].

All of these works deal with the *average* pulse. The ensemble average removes the random interference between the spectral components of the source, and it therefore says nothing about the pulse-to-pulse fluctuations that ultimately set the error rate. Marcuse addressed this question in a fourth paper [7]. The source was modeled as $N$ equally spaced sinusoids of fixed amplitude and random phase, representing a laser oscillating freely in several longitudinal modes. For first-order dispersion he obtained the variance of the detected power in closed form. He showed that the fluctuations disappear on long fibers if the mode spacing exceeds the spectral width of the modulating pulse, because the pulses carried by the individual modes then separate in time. He also argued that this mechanism is ineffective in practical systems in which the pulse broadens by less than a factor of two. With second-order dispersion only, the individual pulses were computed numerically, and the fluctuations were judged to vanish as well because the envelopes of different realizations looked alike [7].

In this paper we generalize the fluctuation analysis of Ref. [7] in the same spirit that Ref. [4] generalized Refs. [1–3]. We include an arbitrary number of dispersion terms, a linearly chirped source, a finite linewidth of each mode and an arbitrary, possibly skewed, distribution of mode powers. The main results are the following. (i) For any linear dispersive medium the mean and variance of the detected power are determined by the single-mode output intensities (Section 3). With first-order dispersion everything is in closed form, including chirp and linewidth (Section 4). (ii) On long fibers the relative fluctuations depend only on the *shape* of the group-delay curve $\tau(x)$, and not on the order or magnitude of the dispersion (Section 5). A monotonic $\tau$ leads to the vanishing fluctuations found by Marcuse. A $\tau$ that is symmetric about a frequency $x_0$ makes pairs of modes placed symmetrically about $x_0$ produce identical intensity profiles, and their beat noise never decays. This is always the case for second-order dispersion alone. It corrects the conclusion drawn from the individual pulses in Ref. [7], in which the random phases only shift fringes that are too fast to be seen on the plots. (iii) Chirp, mode linewidth and spectral skew change the fluctuation level in ways that can be stated quantitatively (Section 6). (iv) With a finite receiver bandwidth (Section 3), dispersion down-converts the beat notes of temporally overlapping modes to baseband, so that the long-fiber fluctuations of a monotonic group delay cannot be filtered out, whereas the persistent pair noise of a symmetric group delay can. (v) The detected power at a given instant is a sum of random phasors (Section 3). Its distribution is far from Gaussian, and the probability of deep fades, which sets the error floor, can be estimated with Gaussian statistics only within orders of magnitude. (vi) Mode-partition noise [8] enters the same formalism in closed form (Section 5). It is absent at the input, grows linearly with the spread of group delays and saturates at a level set by the number of modes. (vii) Marcuse's practical conclusion that fluctuations are unaffected by propagation over links with limited pulse broadening holds for beat noise, but not once mode partition is included.

## 2. Model

We follow the notation of Refs. [4,7]. The unmodulated source consists of $N$ longitudinal modes,

$$\psi_0(t) = \sum_{j=1}^{N} A_j \exp\left[i\left(\omega_j + \delta\omega_j\right)t + i\phi_j\right], \qquad A_j^2 = P_0\, p_j, \quad \sum_j p_j = 1, \quad (1)$$

where the phases $\phi_j$ are independent and uniformly distributed in [0, $2\pi$). Following Ref. [7] the mode amplitudes $A_j$ are fixed. The finite linewidth of each mode is represented by a random frequency offset $\delta\omega_j$, independent from mode to mode, with probability density $\propto \exp(-\delta\omega^2/W_m^2)$. The offset is assumed constant over the duration of a pulse. This quasi-static model describes slow frequency noise, e.g., the $1/f$ and technical frequency noise of semiconductor lasers [9]. Its second-order coherence function, $\langle\psi_{0j}(t)\psi_{0j}^*(t+\tau)\rangle \propto \exp[-(\tau W_m/2)^2]$, is exactly the Gaussian autocorrelation used in Refs. [1,4], so the *mean* pulses are identical to those obtained there. The opposite limit of fast, chaotic mode fields is discussed in Section 3. The mode powers follow Marcuse's Gaussian envelope [7], $p_j \propto \exp\left(-x_j^2/V^2\right)$, or its skewed version [2]. The latter uses widths $V_1 = 2V/\left(1 + R^{2/3}\right)$ for $x_j < 0$ and $V_2 = V_1 R^{2/3}$ for $x_j > 0$, where $R$ is the skew parameter of Ref. [2], and keeps the total power fixed.

The source is power modulated by the Gaussian pulse $s(t) = S\exp[-(t/T)^2]$ and linearly chirped, as in Refs. [3,4]: $\psi(0,t) = s^{1/2}(t)\exp[iCt^2/(2T^2)]\,\psi_0(t)$. The fiber is described by the Taylor expansion of its propagation constant about the central source frequency $\omega_s$, with no truncation [4]:

$$\beta(\omega)z = \beta_s z + \dot{\beta}_s z(\omega - \omega_s) + \Phi(x), \qquad \Phi(x) = \sum_{k\geq 2} D_k\, x^k, \qquad D_k = \frac{\beta_k z}{k!\,T^k}, \quad (2)$$

with the normalized variables $x = (\omega - \omega_s)T$, $x_j = \left(\omega_j - \omega_s\right)T$, $\varepsilon_j = \delta\omega_j T$ and $u = \left(t - \dot{\beta}_s z\right)/T$. Here $D_2$ and $D_3$ coincide with Marcuse's $D$ and $B$. The normalized mode linewidth is $V_m = W_m T$, and the normalized group delay is $\tau(x) = \Phi'(x)$. Each mode carries a replica of the modulated pulse. The output field of a pulse riding on a carrier at normalized frequency $\xi$ is

$$a(u;\xi) = \frac{1}{\sqrt{2\pi(1-iC)}} \int_{-\infty}^{\infty} \exp\left[-\frac{(x-\xi)^2}{2(1-iC)} + iux - i\Phi(x)\right] dx, \tag{3}$$

normalized so that $a(u;\xi) = \exp[i\xi u - (1-iC)u^2/2]$ at $z = 0$. We call its intensity $I(u;\xi) = |a(u;\xi)|^2$, which satisfies $\int I\, du = \sqrt{\pi}$ at every $z$. The detected power, normalized to $SP_0$, is

$$\Pi(u) = \left|\sum_j \sqrt{p_j}\, e^{i\phi_j}\, a(u; x_j + \varepsilon_j)\right|^2. \tag{4}$$

For $C = 0$, $\varepsilon_j = 0$ and $D_k = 0$ for $k \geq 3$, Eq. (4) reduces to Eqs. (17) and (22) of Ref. [7].

Two further ingredients bring the model closer to a real link. First, the mode powers of a multimode laser fluctuate from pulse to pulse while their sum remains nearly constant, which is the origin of mode-partition noise [8]. Following Ogawa, we replace $p_j$ in Eq. (4) by random powers $\pi_j$, independent of the phases, with $\langle \pi_j \rangle = p_j$, $\sum_j \pi_j = 1$ and

$$K_{jk} \equiv \mathrm{Cov}(\pi_j, \pi_k) = k^2 (p_j \delta_{jk} - p_j p_k), \tag{5}$$

where $0 \leq k \leq 1$ is the mode-partition coefficient. The value $k = 0$ gives Marcuse's fixed amplitudes, and $k = 1$ describes a laser that emits in a single, randomly chosen mode in each pulse. Second, the photocurrent is filtered by the receiver. We write the detected signal as $\Pi_h = h * \Pi$, where $h(u)$ is the normalized impulse response. In the examples we use a Gaussian response with transfer function $H_r(X) = \exp(-X^2/B^2)$, where $X$ is the normalized angular frequency conjugate to $u$. The input power pulse $e^{-u^2}$ has the spectrum $\exp(-X^2/4)$, so $B = 2$ corresponds to a receiver matched to the signal.

## 3. Mean and variance for arbitrary dispersion

Averaging Eq. (4) over the phases, only the terms $j = k$ survive in $\langle \Pi \rangle$. In $\langle \Pi^2 \rangle$ the pairings $(j = k, l = m)$ and $(j = m, l = k)$ survive, and the term with all four indices equal is counted once [10]. Averaging then over the independent frequency offsets gives (Appendix A)

$$\langle \Pi(u) \rangle = \sum_j p_j\, \bar{I}_j(u), \tag{6}$$

$$\sigma_\Pi^2(u) = \sum_{j \neq k} p_j\, p_k\, \bar{I}_j(u) \bar{I}_k(u) + \sum_j p_j^2 \left[M_j(u) - \bar{I}_j^2(u)\right], \tag{7}$$

where $\bar{I}_j = \langle I(u; x_j + \varepsilon) \rangle_\varepsilon$ and $M_j = \langle I^2(u; x_j + \varepsilon) \rangle_\varepsilon$. Eqs. (6) and (7) hold for *any* $\Phi(x)$ and for any linear element placed after the modulator. They generalize Eqs. (24) and (25) of Ref. [7], which are recovered for $V_m = 0$ ($M_j = \bar{I}_j^2$) and first-order dispersion. Several conclusions follow immediately.

(1) The first term in Eq. (7) contains only *intensities*. Modes $j$ and $k$ contribute beat noise at time $u$ only if their output intensity profiles overlap there, irrespective of the relative phase of their fields. The problem of fluctuation propagation therefore reduces to that of the single-mode intensities $I(u;\xi)$.

(2) At the input, $\bar{I}_j = e^{-u^2}$ for every mode, and the relative fluctuation $R(u) = \sigma_\Pi / \langle \Pi \rangle$ equals $\rho_{\mathrm{in}} = \left(1 - \sum_j p_j^2\right)^{1/2}$ at all times. $1/\sum_j p_j^2$ is the effective number of modes. For Marcuse's source ($N = 11$, $V = 2\Delta x$) $\rho_{\mathrm{in}} = 0.895$, the value $\approx 0.9$ at $D = 0$ in his Fig. 7.

(3) If instead each mode field is a circular complex Gaussian process (fast chaotic fluctuations, as in amplified spontaneous emission or a light-emitting diode), then $\langle I^2 \rangle = 2\bar{I}^2$ and Eq. (7) gives $\sigma_\Pi = \langle \Pi \rangle$ exactly, for any dispersion, chirp and mode spectrum. This extends Marcuse's LED result, Eq. (27) of Ref. [7], to arbitrary dispersion. In this limit, dispersion cannot reduce the fluctuations at all.

The mean intensity $\bar{I}_j$ is the average pulse of a single-line source of width $V_m$ centered at $x_j$. It is therefore given by Eqs. (12)–(13) of Ref. [4], with the dispersion coefficients re-expanded about $x_j$,

$$D_k^{(j)} = \sum_{m\geq k} \binom{m}{k} D_m x_j^{m-k}, \tag{8}$$

and a delay $u \to u - \tau(x_j)$. Equivalently, its spectral function, $\tilde{I}_j(x) = \int \bar{I}_j(u)e^{-iux}du$, is

$$\tilde{I}_j(x) = \frac{1}{\sqrt{1+C^2}} \int\int S(y+x)S^*(y)\, e^{-i[\Phi(x_j+\varepsilon+y+x)-\Phi(x_j+\varepsilon+y)]}\, f(\varepsilon)\, dy\, d\varepsilon, \tag{9}$$

where $S(y) = \exp[-y^2/2(1-iC)]$ and $f(\varepsilon) = \exp(-\varepsilon^2/V_m^2)/(\sqrt{\pi}V_m)$. The phase difference $\Phi(\cdot+x) - \Phi(\cdot)$ has the polynomial structure of Eq. (7) of Ref. [4]. By Parseval's theorem, the pulse-integrated beat noise follows from the spectral functions alone: $\int \bar{I}_j\bar{I}_k\, du = (2\pi)^{-1}\int \tilde{I}_j\tilde{I}_k^*\, dx$.

To summarize the fluctuations of a whole pulse we use the energy-weighted rms relative fluctuation

$$\rho^2 = \frac{\int \sigma_\Pi^2(u)/\langle\Pi(u)\rangle\, du}{\int \langle\Pi(u)\rangle\, du} = \frac{\int R^2(u)\,\langle\Pi(u)\rangle\, du}{\int \langle\Pi(u)\rangle\, du}, \tag{10}$$

i.e., the mean of $R^2(u)$ weighted by the probability that a photon of the pulse is detected at $u$. At the input $\rho = \rho_{\text{in}}$. Unlike $R(0)$, which Marcuse used, $\rho$ is insensitive to where the pulse is sampled, and, as shown in Section 5, its long-fiber limit has a universal form.

**Receiver bandwidth and mode partition.** With the receiver and the random mode powers, the detected signal is $\Pi_h = \sum_{j,k} (\pi_j\pi_k)^{1/2}\, e^{i(\phi_j-\phi_k)}B_{jk}(u)$, where $B_{jk} = h * (a_j a_k^*)$ and $a_j = a(u; x_j + \varepsilon_j)$. The same phase averaging as in Appendix A gives $\langle\Pi_h\rangle = \sum_j p_j\, \bar{b}_j$ and

$$\sigma_{\Pi_h}^2 = \sum_j \left[(p_j^2 + K_{jj})\, m_j - p_j^2\bar{b}_j^2\right] + \sum_{j\neq k} K_{jk}\; \bar{b}_j\bar{b}_k + \sum_{j\neq k} (p_j p_k + K_{jk}) \left\langle |B_{jk}|^2 \right\rangle_\varepsilon, \tag{11}$$

with $\bar{b}_j = \langle B_{jj}\rangle_\varepsilon$ and $m_j = \langle B_{jj}^2\rangle_\varepsilon$. For $h = \delta$ and $k = 0$ this reduces to Eq. (7). The beat terms now contain the filtered products of *fields*, $h * (a_j a_k^*)$, and are therefore sensitive to the frequency at which two modes beat. For instantaneous detection and $V_m = 0$, Eq. (11) takes the transparent form

$$\sigma_\Pi^2 = (1-k^2)\sum_{j\neq k} p_j\, p_k\, I_j I_k + k^2 \left[\sum_j p_j\, I_j^2 - \left(\sum_j p_j\, I_j\right)^2\right], \tag{12}$$

where $I_j = I(u; x_j)$. Mode partition reduces the beat noise by the factor $1 - k^2$ and adds a term proportional to the variance of the output intensity profiles across the modes. This term vanishes at the input, where all modes carry the same pulse, and it grows as dispersion separates them. Dispersion therefore converts beat noise into mode-partition noise.

**Statistics of the detected power.** For fixed amplitudes and instantaneous detection, $\Pi(u) = \left|\sum_j A_j\, e^{i\phi_j}\right|^2$ with $A_j = [p_j I_j(u)]^{1/2}$ is the squared modulus of a sum of random phasors. Its distribution is given exactly by the Kluyver integral [10]

$$\Pr\{\Pi < y\} = \sqrt{y}\int_0^\infty J_1(q\sqrt{y}) \prod_j J_0(qA_j)dq. \tag{13}$$

For a balanced pair it reduces to the arcsine law $\Pr\{\Pi < y\langle\Pi\rangle\} = \pi^{-1}\arccos(1-y)$, $0 \leq y \leq 2$, which behaves as $(2y)^{1/2}/\pi$ for small $y$. For many modes of comparable power it tends to the exponential law $1 - e^{-y}$. Neither is well described by a Gaussian of the same variance, which predicts a floor $\Phi(-1/R)$ for $\gamma \to 0$. As a pulse-level measure of the error floor we use the energy-weighted fade probability

$$F(\gamma) = \frac{\int \langle \Pi(u)\rangle \Pr\{\Pi(u)<\gamma\langle \Pi(u)\rangle\}\, du}{\int \langle \Pi(u)\rangle\, du}, \tag{14}$$

the probability that the detected power of a "one" falls below a fraction $\gamma$ of its mean. For a noiseless receiver with decision threshold at $\gamma\langle \Pi\rangle$, this is the probability of error on a transmitted one.

## 4. First-order dispersion: closed form

For $\Phi = Dx^2$ the integral in Eq. (3) is elementary: $I(u;\xi) = w^{-1}\exp[-(u-2D\xi)^2/w^2]$, with $w^2 = (1+2DC)^2 + 4D^2$. Averaging over the Gaussian offset gives

$$\bar{I}_j = \frac{1}{W}\exp\left[-\frac{(u-2Dx_j)^2}{W^2}\right], \qquad M_j = \frac{1}{wQ}\exp\left[-\frac{2(u-2Dx_j)^2}{Q^2}\right], \tag{15}$$

with $W^2 = w^2 + 4D^2V_m^2$ and $Q^2 = w^2 + 8D^2V_m^2$. Inserted in Eqs. (6) and (7), these give the mean and variance in closed form for arbitrary chirp, linewidth and mode powers. The rms width of $\langle \Pi\rangle$ reproduces Eq. (26) of Ref. [3] (with $B=0$) for a single mode, and, for a dense comb with a Gaussian envelope, the same expression with $V^2$ replaced by the compound width $V^2 + V_m^2$ [2]. The pulse-integrated overlap of two modes is

$$\frac{\int \bar{I}_j\bar{I}_k\, du}{\int \bar{I}_j^2\, du} = \exp\left[-\Lambda(D)(x_j-x_k)^2\right], \qquad \Lambda(D) = \frac{2D^2}{(1+2DC)^2+4D^2(1+V_m^2)}. \tag{16}$$

Marcuse's separation condition, Eq. (28) of Ref. [7], reads $2\Lambda\Delta x^2 \ge K$. As $D\to\infty$, $\Lambda \to [2(1+C^2+V_m^2)]^{-1}$, so fluctuations can be suppressed only if

$$\Delta x \ge [K(1+C^2+V_m^2)]^{1/2}, \qquad \text{i.e.} \qquad 2T\Delta f \ge \frac{[K(1+C^2+V_m^2)]^{1/2}}{\pi}, \tag{17}$$

which generalizes Marcuse's Eq. (32). Chirp and linewidth broaden the pulse spectrum carried by each mode and demand a proportionally larger mode spacing. For $DC<0$, $\Lambda(D)$ is not monotonic. With $V_m = 0$ it reaches its maximum value $1/2$ exactly at the compression point $D = -1/(2C)$, where the chirp has been compensated. At that point the modes are as well separated as unchirped modes at infinite distance, and their separation then relaxes to the smaller value $1/[2(1+C^2)]$. The jitter term of Eq. (7) vanishes at $D=0$. For an isolated mode its pulse integral, relative to $\int \bar{I}_j^2\, du$, equals $W/w - 1$, which tends to $[1+V_m^2/(1+C^2)]^{1/2} - 1$ on long fibers. The energy-weighted limit is derived for arbitrary dispersion in Section 5.

## 5. Long fibers: stationary-phase analysis

When $|\Phi''| \gg 1$ over the source band, Eq. (3) can be evaluated by stationary phase (Appendix B). This regime is reached on long fibers, and it is the regime of the dispersive Fourier transform [11]. Each stationary point $x_b(u)$, a root of $\tau(x_b) = u$, contributes

$$I(u;\xi) \simeq \sum_b \frac{g(x_b-\xi)}{\sqrt{1+C^2}\,|\Phi''(x_b)|}, \qquad g(y) = \exp\left[-\frac{y^2}{1+C^2}\right], \tag{18}$$

where interference between different branches of a single mode is neglected. It matters only near extrema of $\tau$, i.e., near caustics. Each time instant thus samples the spectrum at the frequencies $x_b(u)$ that are mapped onto it. The sampling window $g$ has width $(1+C^2)^{1/2}$, which is the spectral width of the pulse carried by each mode.

**Monotonic group delay.** If $\tau(x)$ is monotonic over the source band, a single branch exists. The Jacobian $|\Phi''|$ cancels in $R(u)$ and in Eq. (10) after the change of variable $du = |\Phi''|dx$, and

$$\rho_\infty^2 = \frac{1}{\sqrt{\pi(1+C^2)}} \int \frac{\sum_{j\neq k} p_j p_k\, g(x-x_j)\, g(x-x_k)}{\sum_j p_j\, g(x-x_j)}\, dx. \tag{19}$$

This result is *universal*: first-order dispersion, fourth-order Taylor terms, and any combination of dispersion coefficients whose $\Phi''$ keeps a constant sign all lead to the same residual fluctuation. Only the rate at which it is approached differs. For $u$ mapped to a mode frequency, $R(u)$ reproduces the asymptotes of Marcuse's Fig. 7.

**Symmetric group delay and pair degeneracy.** Suppose now that $\Phi(x_0+y) - \Phi(x_0) - \Phi'(x_0)y$ is an odd function of $y$. This is the case for $D_3$ alone ($x_0 = 0$), for any combination of odd-$k$ terms $D_3, D_5, \dots$ ($x_0 = 0$), and for $D_2$ and $D_3$ together with $x_0 = -D_2/(3D_3)$, the zero-dispersion frequency. Changing $y \to -y$ in Eq. (3) gives the exact identity (Appendix C)

$$I(u; x_0 - \eta; C) = I(u; x_0 + \eta; -C) \qquad \text{for all } u \text{ and } z. \tag{20}$$

For an unchirped source, two modes placed symmetrically about $x_0$ therefore have *identical* output intensity profiles at every distance, and by Eq. (7) their beat noise never decays. Where only such a pair $(j, j')$ contributes,

$$R = \frac{(2p_j p_{j'})^{1/2}}{p_j + p_{j'}} \leq \frac{1}{\sqrt{2}}, \tag{21}$$

with equality for balanced pairs. The fields of the two modes carry carriers $x_0 \pm \eta$. Their beat therefore appears as fringes at angular frequency $2\eta/T$ under a common envelope, and a change of the random phases only translates these fringes. This is why the envelopes of Marcuse's Fig. 5 look alike for different phase sets although the power at a fixed instant fluctuates between zero and twice its mean. With two branches of equal $|\Phi''|$ the long-fiber limit becomes

$$\rho_\infty^2 = \frac{\int_0^\infty [\sum_{j\neq k} p_j p_k G_j G_k / \sum_j p_j G_j]\, dy}{\int_0^\infty \sum_j p_j G_j\, dy}, \tag{22}$$

with $G_j(y) = g(x_0 + y - x_j) + g(x_0 - y - x_j)$,

which is again independent of the magnitude and order of the dispersion. For a comb centered on $x_0$ with a mode of power $p_c$ at $x_0$, $\rho_\infty^2 \simeq (1-p_c)/2$ when $\Delta x \gg (1+C^2)^{1/2}$.

**General criterion.** More generally, $\tau$ may take the same value at several frequencies. For each mode $a$, let $\{\hat{x}\}$ be the set of *image frequencies* solving $\tau(\hat{x}) = \tau(x_a)$, with $\hat{x} = x_a$ included. In the long-fiber limit the fluctuations vanish only if no other mode lies within the sampling window of any image,

$$|x_b - \hat{x}| \gtrsim [K(1+C^2)]^{1/2} \qquad \text{for all } b \neq a \text{ and all images } \hat{x} \text{ of } x_a. \tag{23}$$

Marcuse's condition, Eq. (17), is the special case $\hat{x} = x_a$. The additional images appear as soon as $\tau$ is non-monotonic over the source band. For a comb offset by $\delta$ from the nearest symmetric position about $x_0$, the image mismatch of the pairs is $2\delta$, independent of the pair index and of the fiber length. The residual overlap at the peak of a mode is then $\propto \exp[-4\delta^2/(1+C^2)]$.

**Finite linewidth.** With frequency jitter, $g$ in Eq. (18) is convolved with $f(\varepsilon)$. In Eq. (7), $\bar{I}_j$ and $M_j$ are replaced by $\exp\left[-(x-x_j)^2/a^2\right]/a$ and $\exp\left[-2(x-x_j)^2/b^2\right]/(sb)$, respectively, with $s^2 = 1 + C^2$, $a^2 = s^2 + V_m^2$ and $b^2 = s^2 + 2V_m^2$. The cross terms between branches follow in the same way (Appendix B). For an isolated mode the integrals in Eq. (10) are Gaussian and give the exact result

$$\rho_\infty = \frac{V_m}{\sqrt{1+C^2}}. \tag{24}$$

Dispersion converts the frequency jitter into a timing jitter of the pulse carried by each mode. The resulting noise is largest on the pulse slopes, and it persists after the modes have separated, as anticipated qualitatively by Marcuse [7].

**Beat frequencies in the far field.** To the same order, the stationary-phase field of a mode is $a(u;\xi) \propto S(x_b-\xi)\exp\{i[ux_b-\Phi(x_b)]\}$. Its phase depends on $u$ only through $x_b(u)$, which is the same for all modes on a given branch. The product $a_j a_k^*$ of two modes that overlap on the same branch therefore has no carrier. Dispersion down-converts their beat note, which oscillates at $(x_j-x_k)/T$ at the input, to baseband. It cannot then be removed by any receiver whose bandwidth exceeds the inverse duration of the dispersed pulse, $\sim|\Phi''|^{-1}$, and the limit of Eq. (19) is independent of the receiver. By contrast, modes sampled on different branches beat at $(x_{b1}-x_{b2})/T$, i.e., at $2|x-x_0|/T$ for a symmetric group delay. The pair noise of Eq. (21) is therefore removed by a receiver narrower than the pair separation.

**Mode partition.** For isolated modes, Eq. (12) gives $R^2=k^2(1/p_j-1)$ at the arrival time of mode $j$, so the relative fluctuations are largest on the weak side modes. Weighting by energy, the long-fiber limit for a monotonic group delay is

$$\rho_\infty^2=(1-k^2)\rho_{\infty,0}^2+k^2(N-1), \tag{25}$$

where $\rho_{\infty,0}$ is the value of Eq. (19) and $N$ is the number of modes. Each mode contributes $k^2(1-p_j)$ irrespective of its power. In the opposite limit of weak dispersion, the mode pulses are only delayed, $I_j\simeq\exp\left[-(u-\tau_j)^2\right]$ with $\tau_j=\tau(x_j)$. Expanding Eq. (12) to first order in the delays gives

$$\rho_{\mathrm{MPN}}\simeq\sqrt{2}\,k\,\sigma_\tau,\qquad \sigma_\tau^2=\textstyle\sum_j p_j\,\tau_j^2-\left(\sum_j p_j\,\tau_j\right)^2, \tag{26}$$

for any dispersion profile, with $\sigma_\tau=2D_2\sigma_x$ for first-order dispersion, where $\sigma_x$ is the rms width of the mode spectrum. Eq. (26) is linear in the dispersion because the energy weighting includes the pulse slopes, where a delay is most visible. At the pulse center the noise is quadratic in the dispersion, as in Ogawa's analysis [8].

## 6. Results

Unless otherwise stated we use Marcuse's source [7]: $N=11$ modes, $2T\Delta f=1.5$ ($\Delta x=1.5\pi$) and envelope width $V=2\Delta x$. This value of $V$ reproduces his initial fluctuation of 0.9. The mode intensities are computed by fast Fourier transform of Eq. (3) on a common grid. The moments follow from Eqs. (6) and (7), with a 40- to 60-point Gauss–Hermite quadrature over $\varepsilon$ when $V_m>0$. The results were checked against Monte Carlo averages over $2\times10^4$ realizations of Eq. (4).

**Validation.** Fig. 1(a) shows $R(0)$ as a function of $D$ from the closed form of Section 4. It reproduces Fig. 7 of Ref. [7], and the Monte Carlo points (circles) agree with it. Fig. 1(b) tests Eqs. (6) and (7) in a case that combines second-, third- and fourth-order Taylor terms, chirp, mode linewidth and a skewed spectrum. The analytical mean and rms power coincide with the Monte Carlo estimates within their statistical error.

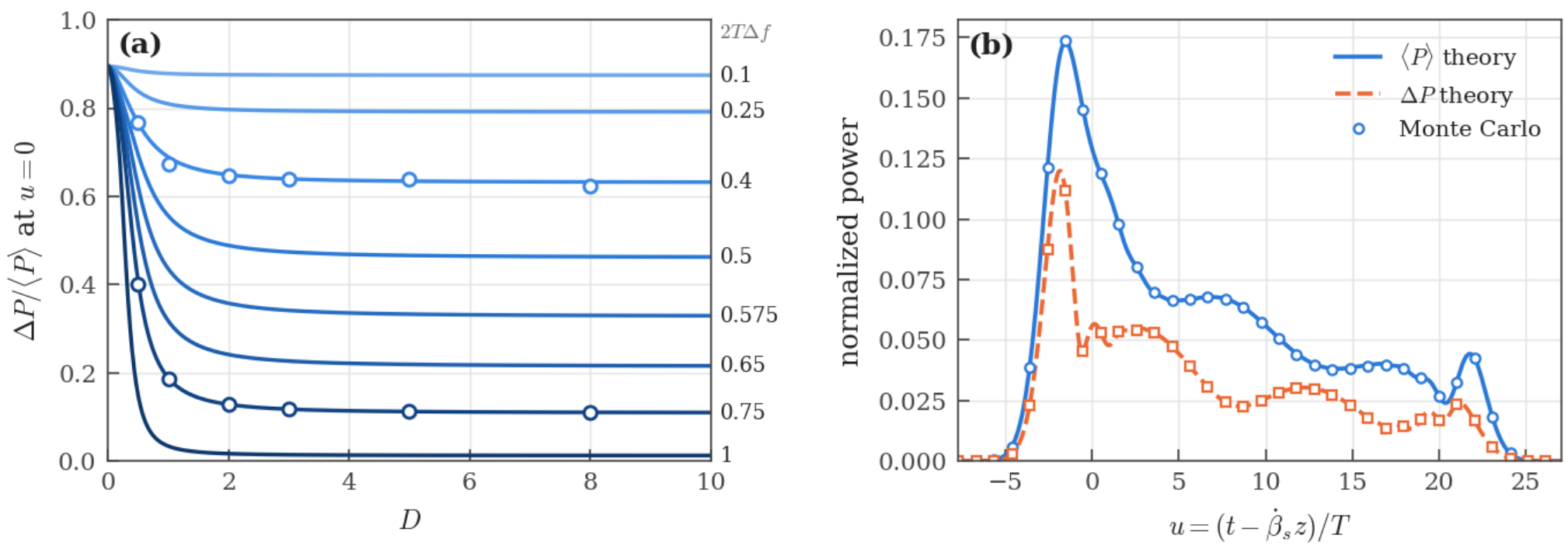

Fig. 1. (a) Relative fluctuation at $u = 0$ as a function of the first-order dispersion parameter $D$ for several values of $2T\Delta f$, the ratio of the mode spacing to the spectral half-width of the pulse (closed form, lines; Monte Carlo, circles). Marcuse's source: $N = 11$, $V = 2\Delta x$. Compare with Fig. 7 of Ref. [7]. (b) Mean and rms power for $N = 7$, $\Delta x = 2$, $V = 3$, $R = 4$, $D_2 = 1.5$, $D_3 = 0.2$, $D_4 = -0.02$, $C = -0.6$, $V_m = 0.3$: theory (lines) and $2 \times 10^4$ Monte Carlo realizations (markers).

**Second-order dispersion alone.** Fig. 2 revisits Fig. 5(c) of Ref. [7] ($D_2 = 0$, $B = D_3 = 0.2$). The central mode arrives first, followed by the pulses carried by the pairs $\pm\Delta x$ and $\pm 2\Delta x$ [Fig. 2(a)]. Each pair pulse is filled with fringes of period $\pi/(n\Delta x)$. A magnified view [Fig. 2(b)] shows that different realizations shift these fringes but do not remove them. The relative fluctuation [Fig. 2(c)] is $1/\sqrt{2}$ throughout the pair pulses, as predicted by Eq. (21), and it drops to zero only on the central mode. For comparison, a first-order dispersion producing the same rms width leaves $R \approx 0$ at the center of each mode pulse, and $R$ is significant only where neighboring modes overlap. The fluctuations had therefore not vanished in Marcuse's Fig. 5. They had become too fast to be resolved on the scale of the plot.

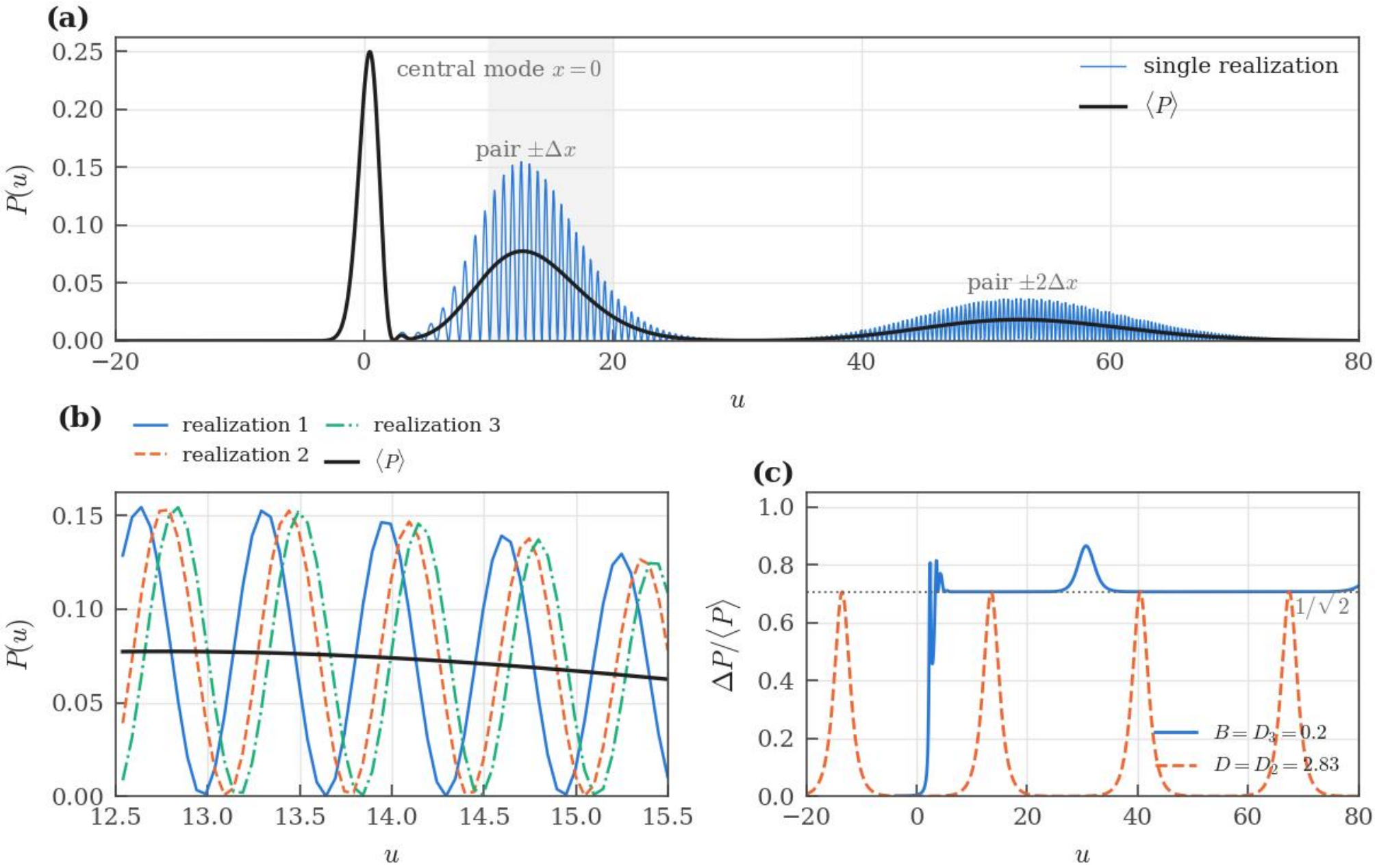

Fig. 2. Second-order dispersion alone, $D_3 = 0.2$, $2T\Delta f = 1.5$, $N = 11$ [conditions of Fig. 5(c) of Ref. [7]]. (a) One realization and the mean pulse. (b) Magnified view of the shaded region for three sets of random phases. (c) Relative fluctuation $R(u)$, compared with that obtained with first-order dispersion alone for the same rms output width.

**Stand-alone dispersion orders.** Fig. 3 shows $\rho$ as a function of the rms broadening of the mean pulse for each Taylor term $D_k$ acting alone ($k = 2, \dots, 7$), the fluctuation counterpart of Fig. 2 of Ref. [4]. The even-$k$ terms ($D_2$, $D_4$, $D_6$) give a monotonic group delay, and they all tend to the same limit, $\rho_\infty = 0.035$ from Eq. (19). The odd-$k$ terms ($D_3$, $D_5$, $D_7$) give a symmetric group delay, and they all tend to $\rho_\infty = 0.600$ from Eq. (22), close to $[(1 - p_c)/2]^{1/2} = 0.599$. The approach is slower for higher orders, because the group-delay curve flattens near $x = 0$ and the central modes separate late. Even at $\sigma/\sigma_0 = 10^3$, $D_6$ and $D_7$ have not reached their limits.

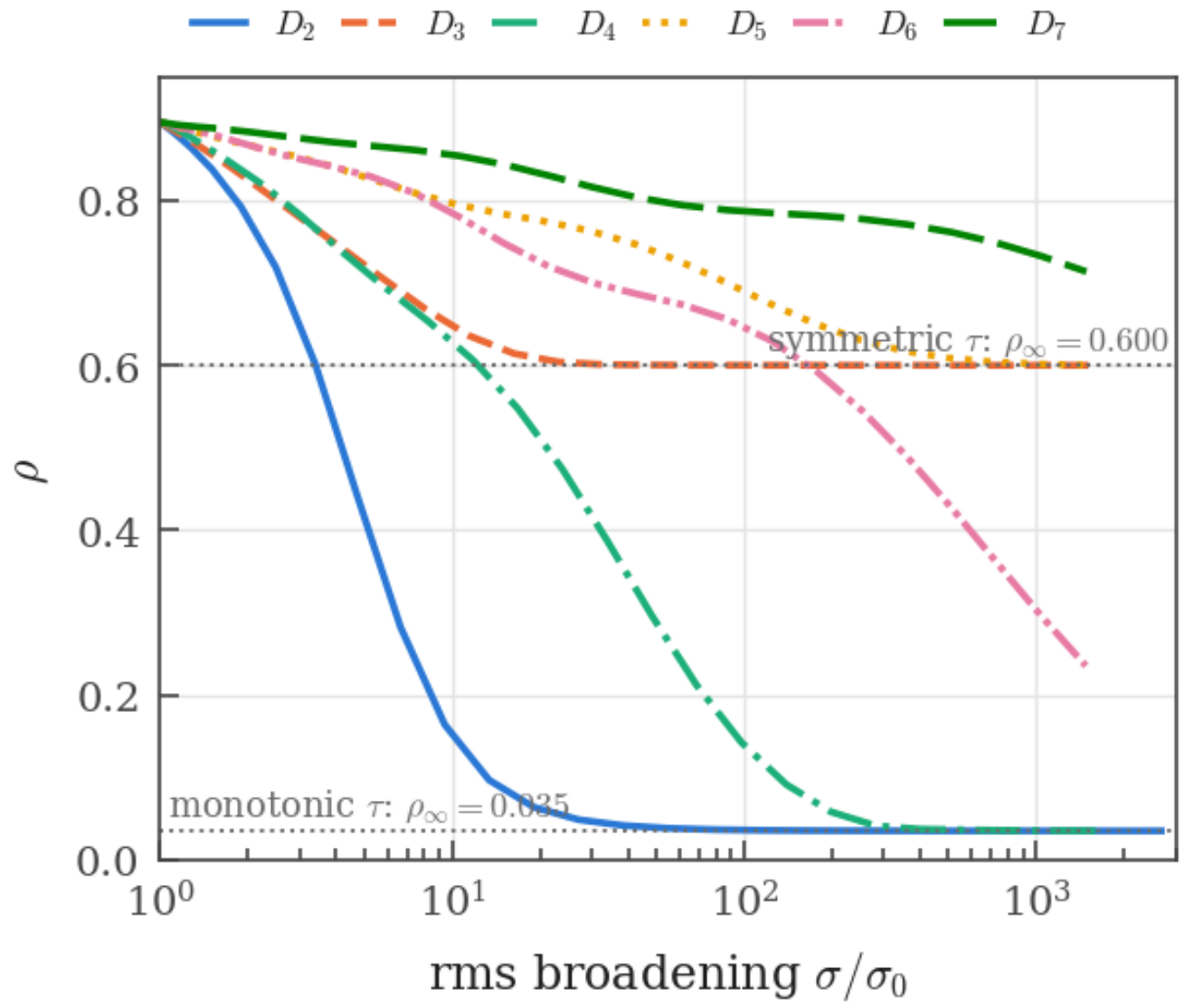

Fig. 3. Energy-weighted relative fluctuation $\rho$ as a function of the rms broadening of the mean pulse, for stand-alone Taylor terms $D_k$. Dotted lines: long-fiber limits, Eq. (19) (monotonic $\tau$) and Eq. (22) (symmetric $\tau$). Source as in Fig. 2.

**Operation near the zero-dispersion wavelength.** With $D_2$ and $D_3$ the symmetry frequency is $x_0 = -D_2/(3D_3)$, and its position relative to the comb controls the long-fiber fluctuation [Fig. 4(a)]. Sharp maxima occur when $x_0$ coincides with a mode ($x_0/\Delta x$ an integer) or lies midway between two modes (half-integer). All pairs are then exactly degenerate. Between these values the image mismatch $2\delta$ reduces the fluctuation, but it remains far above the monotonic limit until $x_0$ leaves the populated part of the spectrum ($x_0 \gtrsim 3$–$4\Delta x$ here). Fig. 4(b) shows the evolution with distance. This is the fluctuation counterpart of Marcuse's finding that the rms width depends on the spectral skew only near $\lambda_0$ [2].

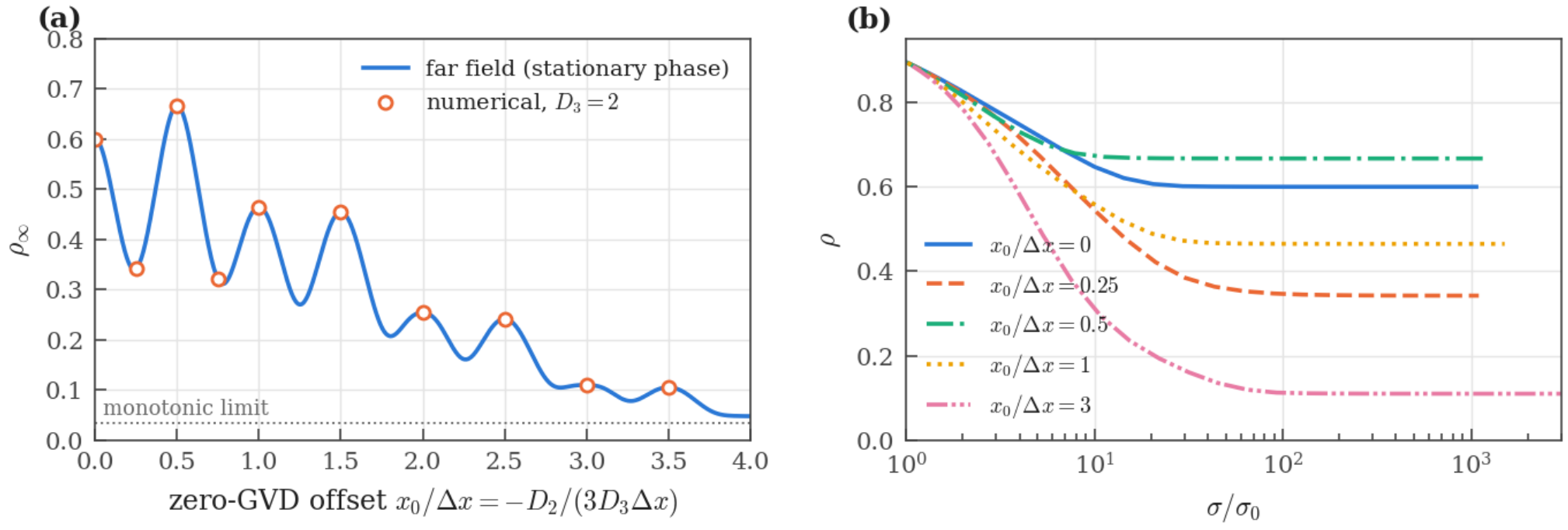

Fig. 4. First- and second-order dispersion. (a) Long-fiber fluctuation as a function of the position of the zero-dispersion frequency $x_0 = -D_2/(3D_3)$ relative to the mode comb: Eq. (22) (line)

and full numerical evaluation at $D_3 = 2$ (circles). (b) $\rho$ as a function of the rms broadening for several $x_0$.

**Chirp.** Fig. 5(a) shows $\rho(D)$ for first-order dispersion and $C = 0, \pm 1, \pm 2$. For $DC < 0$ the fluctuations first drop faster than for $C = 0$. Around the compression point $D = -1/(2C)$ (triangles), $\Lambda$ reaches the unchirped asymptotic value, and the fluctuations then rise to a higher limit. For $DC > 0$ they decay more slowly. Both signs end at the same limit, which increases with $|C|$ because the pulse spectrum carried by each mode is $(1 + C^2)^{1/2}$ times wider [Fig. 5(b)]. With a symmetric group delay, chirp breaks the pair identity only at finite distance [Eq. (20)]. In the far field, $I$ depends on $C$ only through $C^2$, and the pair degeneracy is restored, so chirp cannot remove it.

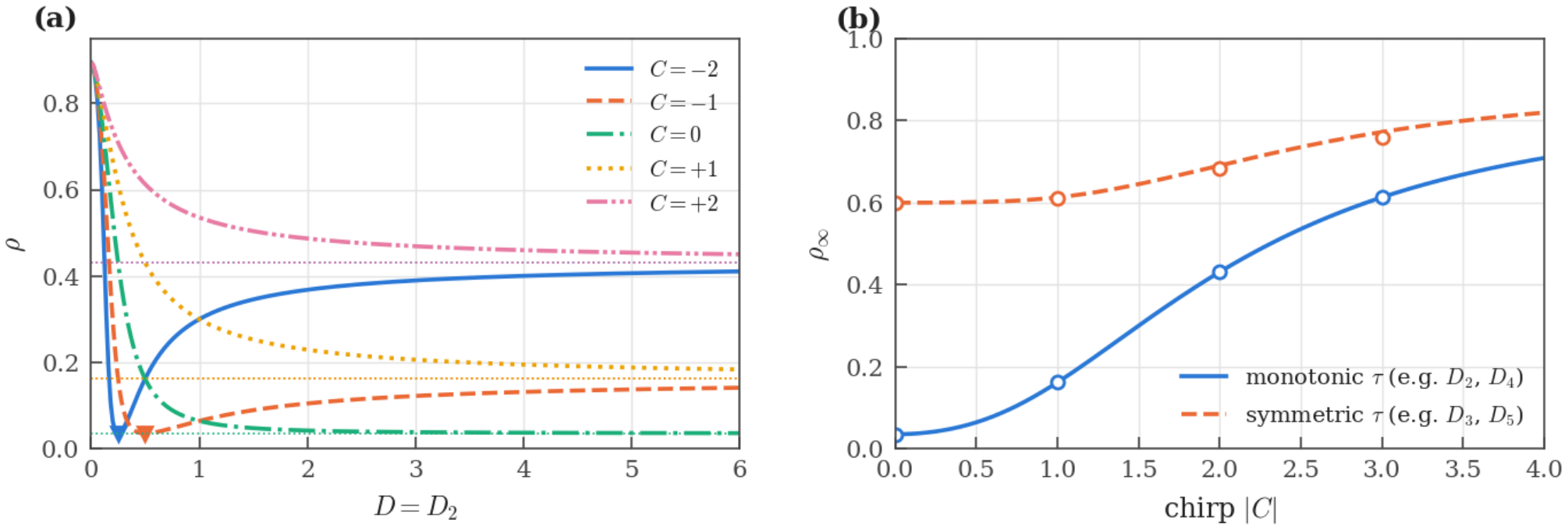

Fig. 5. Source chirp. (a) $\rho$ as a function of $D$ (first-order dispersion) for several values of $C$; dotted lines, long-fiber limits; triangles, compression points $D = -1/(2C)$. (b) Long-fiber limit as a function of $|C|$ for monotonic and symmetric group delay (lines), with numerical results at large dispersion (circles).

**Mode linewidth.** Frequency jitter of the modes sets a floor that grows almost linearly with $V_m$ [Fig. 6]. For monotonic $\tau$ the limit is well approximated by $\left(\rho_{\infty,0}^2 + V_m^2\right)^{1/2}$, where $\rho_{\infty,0}$ is the value for $V_m = 0$, in agreement with Eq. (24). For the symmetric case the jitter adds to the pair beat noise. Its effect becomes dominant only for $V_m \gtrsim 0.5$. The dotted line in Fig. 6(a) marks the chaotic-mode limit $\rho = 1$ of Section 3. For a real laser, whose modes combine slow frequency noise with fast phase diffusion, the fluctuations lie between the two models.

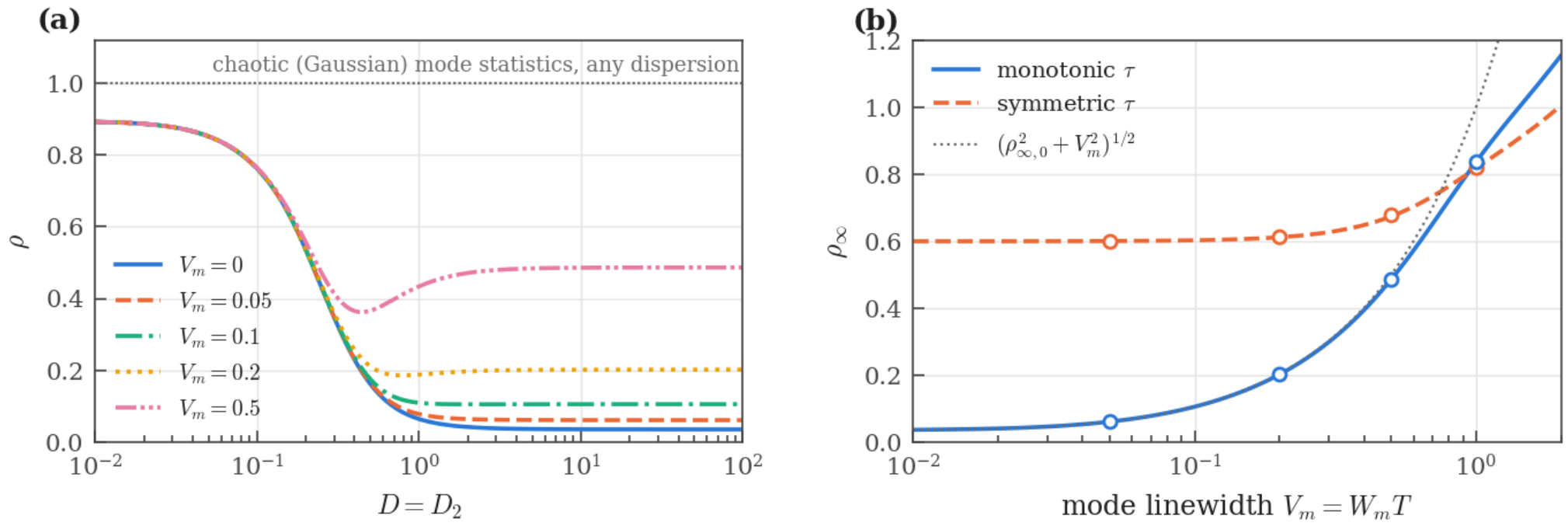

Fig. 6. Finite mode linewidth (quasi-static frequency jitter). (a) $\rho$ as a function of $D$ for several $V_m$ (closed form, Section 4). (b) Long-fiber limit as a function of $V_m$ for monotonic and symmetric group delay (lines), with numerical results (circles).

**Skewed mode spectrum.** Fig. 7 shows the long-fiber fluctuation as a function of Marcuse's skew parameter $R$ [2]. For monotonic $\tau$ the result hardly depends on $R$. For $x_0 = 0$ it is symmetric in $\ln R$, because skewing the spectrum unbalances the pairs and lowers Eq. (21). When $x_0 \neq 0$ lies within the source band, the dependence becomes strongly asymmetric.

Skewing the spectrum towards $x_0$ populates the degenerate pairs and raises the fluctuations, whereas skewing it away suppresses them. For example, at $x_0 = +2\Delta x$, $\rho_\infty$ rises from 0.03 for $R = 0.1$ to 0.56 for $R = 10$.

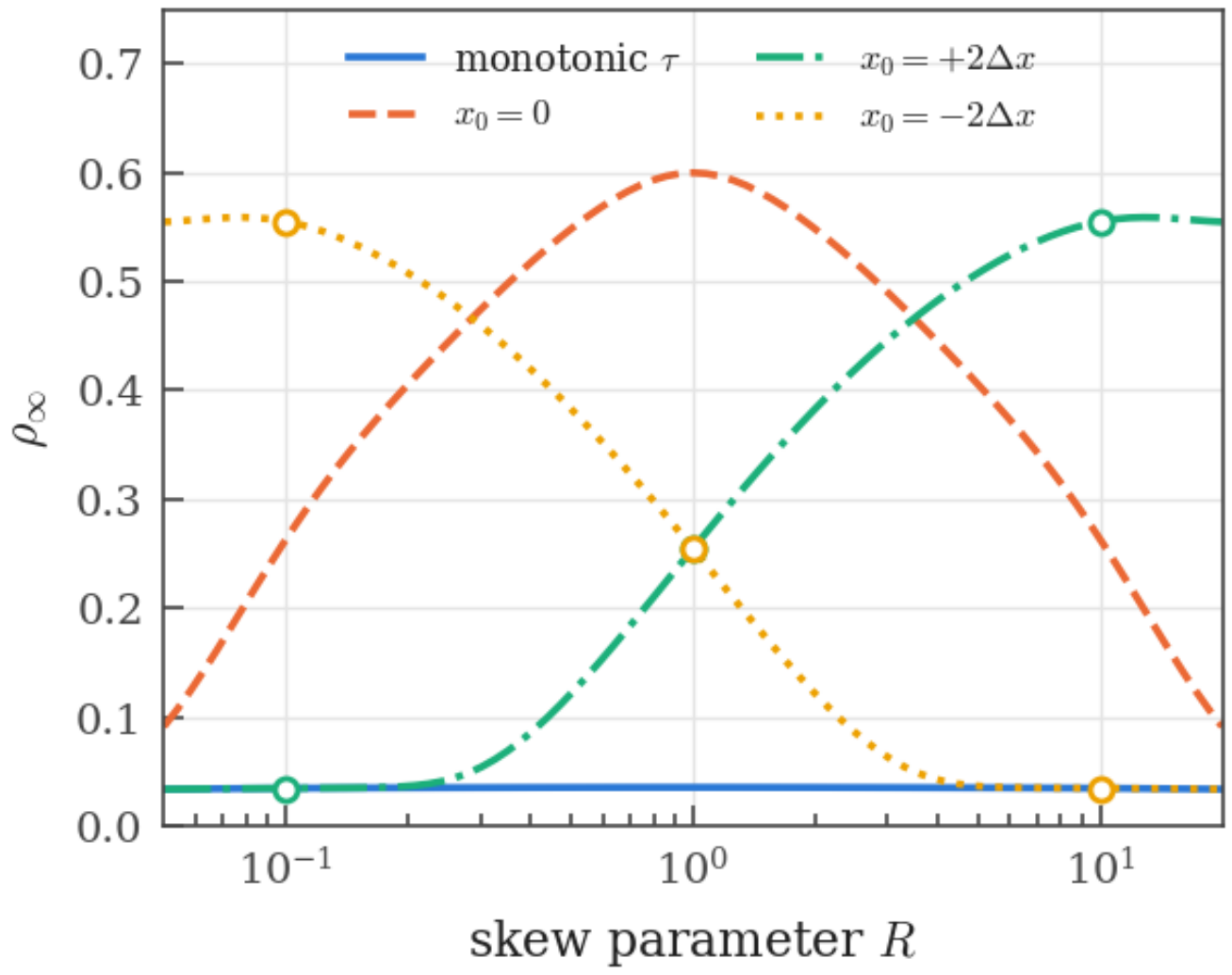

Fig. 7. Long-fiber fluctuation as a function of the skew parameter $R$ of the mode envelope [2], for monotonic group delay and for symmetric group delay centered at $x_0 = 0, \pm 2\Delta x$. Circles: full numerical evaluation at $D_3 = 2$.

**Receiver bandwidth.** Fig. 8(a) shows $\rho$ for Marcuse's source as a function of the receiver bandwidth $B$. At the input, the beat notes lie at multiples of the mode spacing, $\Delta x = 4.7$. A matched receiver ($B = 2$) therefore reduces the input fluctuation from 0.895 to 0.038, and the full value is recovered only for $B \gg \Delta x$. With second-order dispersion alone ($D_3 = 2$) the persistent pair noise, $\rho = 0.60$ for instantaneous detection, is filtered to 0.025 by a matched receiver and reappears when $B$ exceeds the pair beat frequencies $2n\Delta x$. With first-order dispersion ($D_2 = 30$), in contrast, $\rho = 0.035$ for every $B$, because the residual beats have been down-converted to baseband. The practically relevant question is therefore what happens when the mode spacing is comparable to the signal bandwidth, so that the beats cannot be filtered at the input [Fig. 8(b)]. For $2T\Delta f \lesssim 1$ and a matched receiver, first-order dispersion *increases* the fluctuations, for instance from 0.46 to 0.59 for $2T\Delta f = 0.5$ and from 0.69 to 0.79 for $2T\Delta f = 0.25$. The input beats that the receiver partly averaged out are brought to baseband by the dispersion. Second-order dispersion leaves the fluctuations nearly unchanged in this regime.

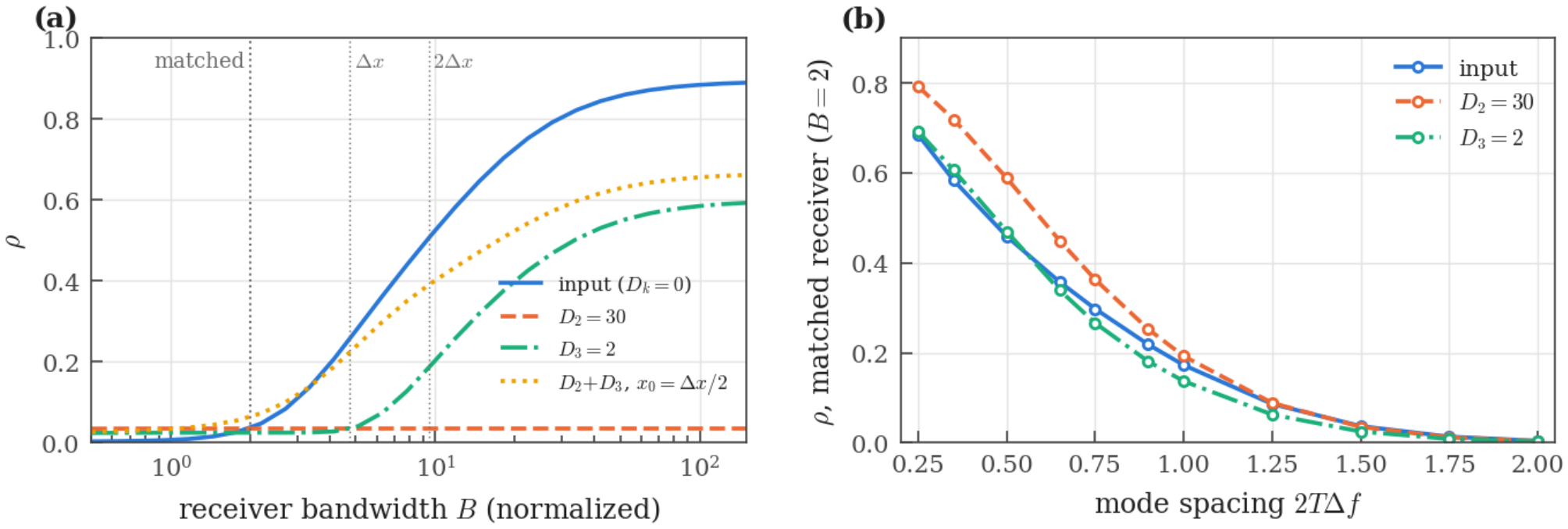

Fig. 8. Receiver bandwidth. (a) Energy-weighted fluctuation $\rho$ as a function of the normalized bandwidth $B$ of a Gaussian receiver, at the input and on long fibers with first-order dispersion, second-order dispersion, and both with $x_0 = \Delta x/2$ (source as in Fig. 2). (b) Matched receiver ($B = 2$): $\rho$ as a function of the mode spacing at the input and on long fibers ($N = 11$, $V = 2\Delta x$).

**Statistics and fade probability.** Fig. 9(a) compares the exact distribution of the detected power, Eq. (13), with the Gaussian distribution of the same variance for a balanced pair and for Marcuse's 11-mode input. The pair follows the arcsine law, whose probability of deep fades decays only as $y^{1/2}$. The 11-mode sum is close to exponential. In both cases the Gaussian approximation predicts a spurious floor of 8% and 13% for $y \to 0$. Fig. 9(b) shows the energy-weighted fade probability of Eq. (14), obtained by Monte Carlo simulation (solid), and its Gaussian estimate (dashed). At $\gamma = 0.01$ the exact values are $9 \times 10^{-3}$ at the input, $3.2 \times 10^{-2}$ for the persistent pair noise of second-order dispersion, $3.5 \times 10^{-3}$ for the far field of a dense comb with first-order dispersion, and $2 \times 10^{-5}$ for mode-partition noise seen through a matched receiver. The corresponding Gaussian estimates are 0.13, 0.058, 0.046 and $9 \times 10^{-3}$. The Gaussian approximation thus overestimates deep fades by up to three orders of magnitude, yet underestimates moderate fades of the pair noise ($F(0.5) = 0.24$ versus 0.17). Error-floor estimates for multimode sources must therefore use the random-phasor statistics rather than the variance alone.

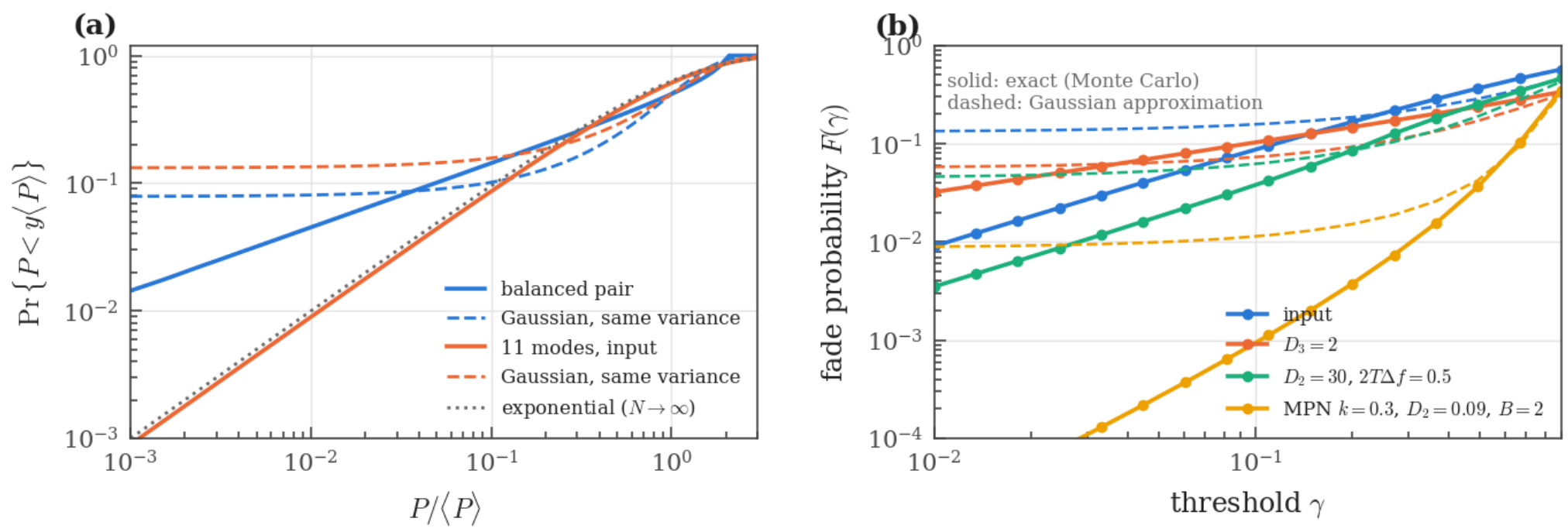

Fig. 9. Statistics of the detected power. (a) Exact cumulative distribution, Eq. (13), for a balanced pair of modes and for the 11-mode input, compared with Gaussian distributions of equal variance and with the exponential limit. (b) Energy-weighted fade probability $F(\gamma)$, Eq. (14), from $2 \times 10^4$ Monte Carlo realizations (solid) and from the Gaussian approximation (dashed): input; second-order dispersion alone ($D_3 = 2$); dense comb ($2T\Delta f = 0.5$) with $D_2 = 30$; mode-partition noise with $k = 0.3$, $D_2 = 0.09$ ($\sigma/\sigma_0 \approx 2$) and a matched receiver.

**Mode-partition noise.** Fig. 10 shows $\rho$ as a function of the rms broadening for first-order dispersion and partition coefficients $k = 0$–$0.6$. For instantaneous detection [Fig. 10(a)] the decay of the beat noise and the growth of the partition noise partly compensate each other. For $k = 0.3$ the total fluctuation changes little along the fiber (0.85 at the input, 0.88 at $\sigma/\sigma_0 = 2$, 0.95 on long fibers), although its nature changes completely. With a matched receiver [Fig. 10(b)] the beat noise is removed and the partition noise dominates. It grows from 0.035 at the input to 0.16, 0.46 and 0.92 at $\sigma/\sigma_0 = 2$ for $k = 0.1$, 0.3 and 0.6, and it saturates at the values of Eq. (25) (dotted lines). The initial growth follows Eq. (26), which we verified numerically to better than 1% for $\rho_{\mathrm{MPN}} < 0.2$.

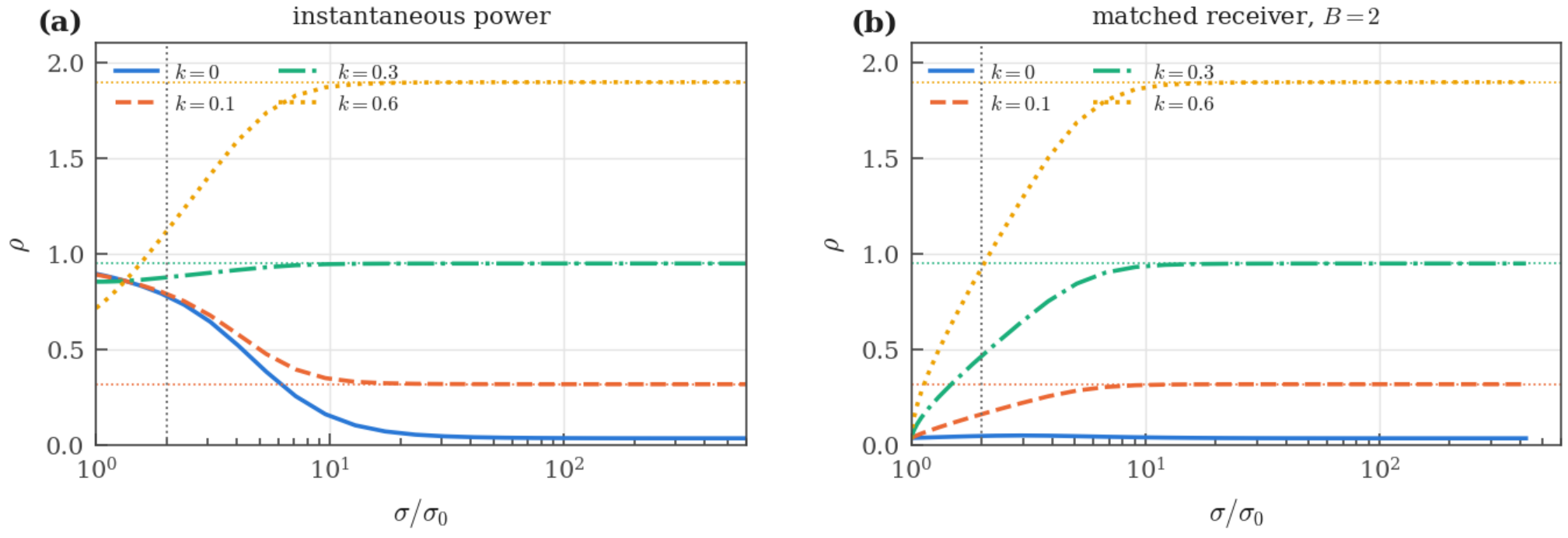

Fig. 10. Mode-partition noise with first-order dispersion, Marcuse's source, and partition coefficients $k = 0$, 0.1, 0.3 and 0.6. (a) Instantaneous detection. (b) Matched receiver ($B = 2$); $\sigma_0$ is the filtered input width. Dotted horizontal lines: long-fiber limits, Eq. (25); vertical line: $\sigma/\sigma_0 = 2$.

**Practical systems.** Marcuse showed that for first-order dispersion and an envelope containing at least a few modes ($V/\Delta x = M \geq 1$), fluctuations cannot decrease if the output pulse is to be at most twice as wide as the input pulse [7]. Fig. 11 tests this conclusion for random dispersion profiles containing Taylor terms $D_2$ to $D_6$ with random signs and magnitudes, random mode spacings $0.5 \leq 2T\Delta f \leq 3$, and three envelope widths. For $\sigma/\sigma_0 \leq 2$, $\rho/\rho_{\text{in}}$ never falls below 0.90 for $M = 2$ or below 0.83 for $M = 1$, which confirms Marcuse's conclusion for arbitrary dispersion. For a nearly single-mode laser ($M = 0.6$, effective number of modes $1/\sum_j p_j^2 \approx 1.25$), reductions to $\rho/\rho_{\text{in}} \approx 0.65$ occur already at $\sigma/\sigma_0 < 2$. The input fluctuation is, however, smaller to begin with in that case ($\rho_{\text{in}} = 0.45$). Larger reductions become possible for broadenings well beyond a factor of two, mainly for envelopes with $M \leq 1$. This confirmation, however, refers to the beat noise alone. As shown by Figs. 8 and 10, a matched receiver removes most of the beat noise of widely spaced modes to begin with, and mode partition makes the fluctuations grow along the link. For $k = 0.3$ and $\sigma/\sigma_0 = 2$ they rise more than tenfold. In practical systems with band-limited receivers, dispersion does not leave the fluctuations unchanged, as Marcuse concluded. It converts beat noise, which the receiver can partly reject, into partition noise, which it cannot.

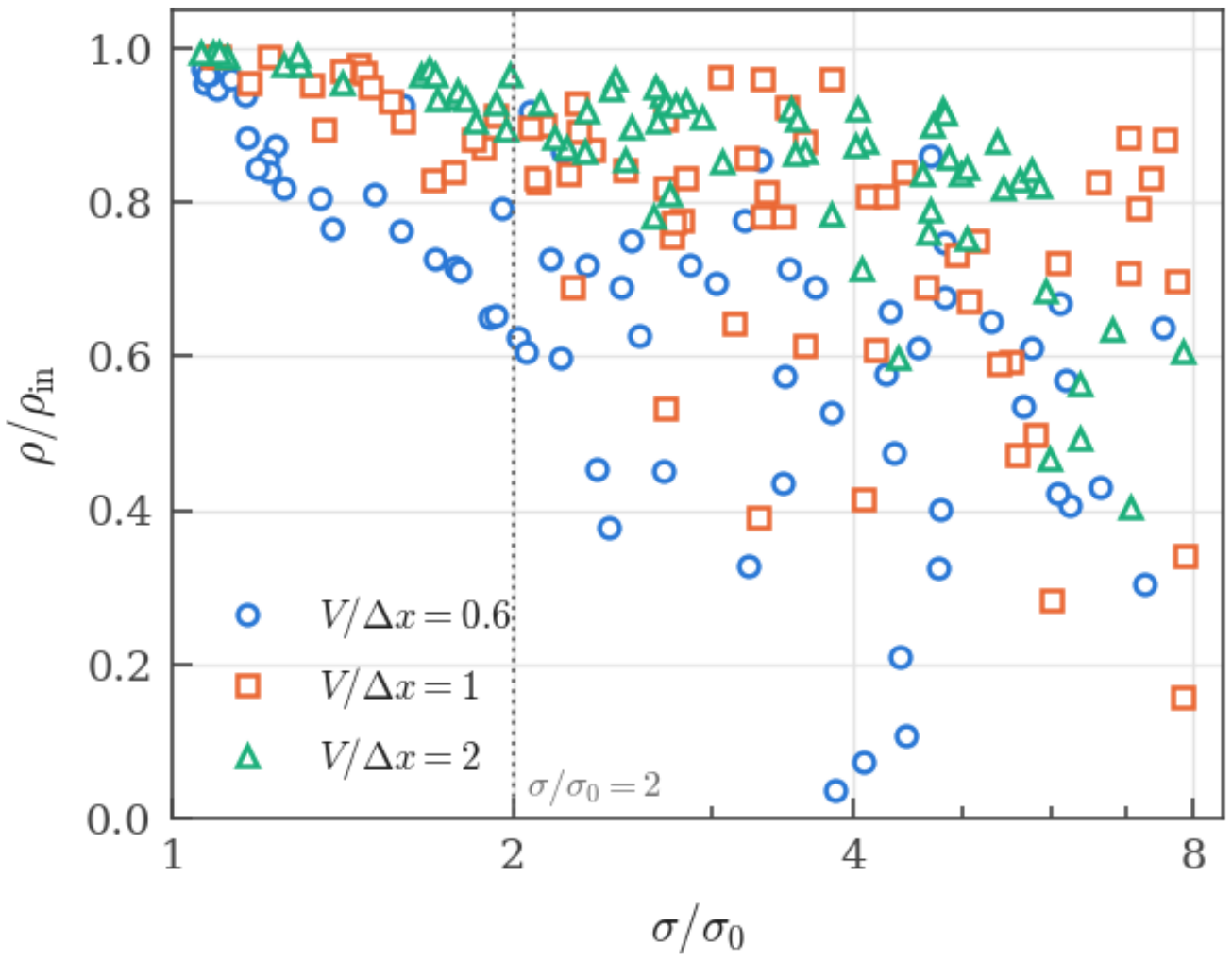

Fig. 11. Fluctuation reduction $\rho/\rho_{\text{in}}$ as a function of the rms broadening for random dispersion profiles ($D_2$ to $D_6$), random mode spacings $0.5 \leq 2T\Delta f \leq 3$ and envelope widths $V/\Delta x =$ 0.6,1,2 ($N = 11$).

## 7. Discussion

The analysis combines three mechanisms that previous treatments considered separately: the beat noise of randomly phased modes [7], mode-partition noise [8], and the conversion of laser frequency noise into intensity noise by dispersion [12]. Eq. (11) contains all three for arbitrary dispersion and receiver response, because only the single-mode output fields enter. Their relative importance depends on the receiver. For instantaneous detection, which is relevant to dispersive Fourier transform and time-stretch measurements with wideband photodetectors [11], the beat noise and its group-delay-dependent evolution dominate, including the persistent pair noise of a symmetric group delay. For a receiver matched to the signal, the beat noise matters only when the mode spacing is comparable to the signal bandwidth, and mode partition and frequency jitter dominate. The distribution of the detected power is in all cases far from Gaussian, so error-rate estimates should be based on Eqs. (13) and (14) rather than on the variance. A laser with Lorentzian (phase-diffusion) linewidth and correlated amplitude and partition fluctuations, as described by rate equations, can be included through its fourth-order field statistics and is the natural next step.

The universality of the long-fiber limits has a simple interpretation in the language of the dispersive Fourier transform and of space–time duality [6,11]. On long fibers the output pulse is a scaled image of the source spectrum, observed through a window whose width equals the spectral width of the modulating pulse. Dispersion only sets the scale of the image and, through $\tau(x)$, which parts of the spectrum are superposed at a given time. Fluctuations therefore disappear when each window contains a single mode, and they persist when the group-delay map folds the spectrum onto itself.

## 8. Conclusions

We have extended Marcuse's analysis of pulse fluctuations in single-mode fibers [7] to arbitrary chromatic dispersion, chirped sources, modes of finite linewidth and skewed mode spectra, in the same way that Ref. [4] extended his analysis of the average pulse. The mean and variance of the detected power are determined by single-mode output intensities, which are available in closed form for first-order dispersion and as a single Fourier integral otherwise. On long fibers the fluctuation level depends only on the topology of the group-delay curve. A monotonic curve suppresses the fluctuations, as Marcuse found for first-order dispersion. A symmetric curve, such as second-order dispersion alone or operation near the zero-dispersion wavelength, leaves pairs of modes with identical intensity profiles and a persistent beat noise of relative magnitude up to $1/\sqrt{2}$. Chirp and mode linewidth increase the residual fluctuations, with closed-form criteria [Eqs. (17) and (24)]. A skewed spectrum makes the fluctuations depend on the direction of the skew only near the zero-dispersion wavelength. The receiver bandwidth distinguishes the two long-fiber regimes: dispersion down-converts the beats of modes that overlap on the same branch of the group-delay curve to baseband, whereas symmetric pairs keep beating at high frequency. The detected power follows random-phasor statistics, and Gaussian estimates of the fade probability can be wrong by orders of magnitude in either direction. Mode-partition noise grows with dispersion as $\sqrt{2}k\sigma_\tau$ and saturates at $k\sqrt{N-1}$. Marcuse's conclusion that fluctuations are practically unaffected in systems with limited pulse broadening holds for the beat noise of sources with several significant modes, but with a band-limited receiver and mode partition the fluctuations grow substantially along the link.

## Appendix A. Fourth-order moment

Write $c_j = \sqrt{p_j}\, a(u; x_j + \varepsilon_j)$. For independent uniform phases, $\left\langle e^{i(\phi_j - \phi_k + \phi_l - \phi_m)} \right\rangle_\phi$ is unity when $(j = k,\, l = m)$ or $(j = m,\, l = k)$ and zero otherwise, so $\left\langle \left|\sum_j c_j\, e^{i\phi_j}\right|^4 \right\rangle_\phi = 2\left(\sum_j |c_j|^2\right)^2 - \sum_j |c_j|^4$. Averaging over the independent $\varepsilon_j$, $\langle \Pi^2 \rangle = 2\sum_{j\neq k} p_j\, p_k \bar{I}_j \bar{I}_k + \sum_j p_j^2\, M_j$. Subtracting $\langle \Pi \rangle^2 = \sum_j p_j^2\, \bar{I}_j^2 + \sum_{j\neq k} p_j\, p_k \bar{I}_j \bar{I}_k$ gives Eq. (7). For circular complex Gaussian mode fields, the Gaussian moment theorem [10] gives $M_j = 2\bar{I}_j^2$ and $\sigma_\Pi^2 = \langle \Pi \rangle^2$.

With random powers and a receiver, the detected signal is the quadratic form $\Pi_h = \sum_{j,k} z_j\, z_k^* B_{jk}$ with $z_j = \pi_j^{1/2} e^{i\phi_j}$ and $B_{kj} = B_{jk}^*$ for real $h$. The phase average gives $\langle \Pi_h \rangle_\phi = \sum_j \pi_j\, B_{jj}$ and $\langle \Pi_h^2 \rangle_\phi = \left(\sum_j \pi_j\, B_{jj}\right)^2 + \sum_{j\neq k} \pi_j\, \pi_k \left|B_{jk}\right|^2$. Averaging over the independent powers and frequency offsets with $\langle \pi_j \pi_k \rangle = p_j p_k + K_{jk}$ and subtracting $\langle \Pi_h \rangle^2$ gives Eq. (11). Inserting Eq. (5) with $h = \delta$ and $V_m = 0$ yields Eq. (12). In the Monte Carlo simulations, Eq. (5) is realized exactly by a Dirichlet distribution of the mode powers with concentration parameter $1/k^2 - 1$.

## Appendix B. Stationary-phase evaluation

With $x = x_b + y$, the phase of the integrand of Eq. (3) near a root of $\tau(x_b) = u$ is $ux_b - \Phi(x_b) - \frac{1}{2}\Phi''(x_b)y^2 + O(y^3)$. When $|\Phi''| \gg 1 + C^2$ the Gaussian factor varies slowly over the stationary region and can be taken outside the integral. This gives $|a|^2 \simeq |S(x_b - \xi)|^2/\left[\sqrt{1+C^2}\,|\Phi''(x_b)|\right]$ per branch and hence Eq. (18). The condition fails near extrema of $\tau$, where neighboring stationary points coalesce into an Airy-type caustic of width $\sim |\Phi'''|^{1/3}$. Its contribution to $\rho$ vanishes relative to the rest of the pulse as $z \to \infty$. With jitter, $\langle g(\alpha - \varepsilon)g(\beta - \varepsilon)\rangle_\varepsilon = (s/b)\exp[-(\alpha - \beta)^2/(2s^2)]\exp[-(\alpha + \beta)^2/(2b^2)]$, with $s$ and $b$ as defined in Section 5. This expression gives $\bar{I}_j$, $M_j$ and the inter-branch terms in closed form. For a single branch, $\int \left(M_j - \bar{I}_j^2\right)/\bar{I}_j\, dx / \int \bar{I}_j\, dx = a^2/s^2 - 1$, which is Eq. (24).

## Appendix C. Symmetry identity

Let $\Phi(x_0 + y) = \Phi_0 + \tau_0 y + \Psi(y)$ with $\Psi$ odd, and write $\xi = x_0 + \eta$. From Eq. (3), $a(u; x_0 + \eta; C) = K_C e^{i(ux_0 - \Phi_0)} \int S_C(y - \eta) e^{i(u - \tau_0)y - i\Psi(y)} dy$, with $K_C = [2\pi(1 - iC)]^{-1/2}$. Replacing $\eta \to -\eta$ and $y \to -y$, and using that $S_C$ is even and $\Psi$ odd, gives $a(u; x_0 - \eta; C) = K_C e^{i(ux_0 - \Phi_0)} \left[\int S_{-C}(y - \eta) e^{i(u - \tau_0)y - i\Psi(y)} dy\right]^*$. Since $|K_C| = |K_{-C}|$, Eq. (20) follows.

**Funding.** This work is not subject to funding. It has been developed and written by the author as a tribute to Dr. Dietric Marcuse pioneering work.

**Disclosures.** The author declares no conflicts of interest.

**Data availability.** The Python code that implements the model and generates all figures, together with the data underlying the figures, is available from the author upon reasonable request.